\documentclass[12pt]{iopart}
\usepackage{graphicx}
\usepackage{subfig}

\newcommand{\beq}{\begin{equation}}
\newcommand{\eeq}{\end{equation}}
\newcommand{\ba}{\begin{eqnarray}}
\newcommand{\ea}{\end{eqnarray}}

\usepackage[english]{babel}
\begin{document}

\title{Clamped seismic metamaterials: Ultra-low broad frequency stop-bands}
\author{Y. Achaoui$^1$,  T. Antonakakis$^2$, S. Br\^ul\'e$^3$,
  R.~V. Craster$^4$, 
    S. Enoch$^1$ and S. Guenneau$^1$}

\address{$^1$ Aix-Marseille Univ., CNRS, Centrale Marseille, Institut Fresnel, 13013 Marseille, France}
\address{$^2$ Multiwave Technologies AG, 3 Chemin du Pr\'e Fleurim 1228,
  Geneva, Switzerland}
\address{$^3$ M\'enard, 91620 Nozay, France}
\address{$^4$ Department of Mathematics, Imperial College London,
  London SW7 2AZ, UK}

\date{\today}

\begin{abstract}
The regularity of earthquakes, their destructive power, and
the nuisance of ground vibration in urban environments, all motivate
designs of defence structures to lessen the impact of seismic
and ground vibration waves on buildings. Low frequency waves, in the range $1$ to
$10$ Hz for earthquakes and up to a few tens of Hz for vibrations
generated by human activities, cause a large amount of damage, or
inconvenience; depending on the geological conditions they can travel considerable distances and may match the resonant
fundamental frequency of buildings. The ultimate aim of any seismic
metamaterial, or any other seismic shield, is to protect over this entire range of
frequencies; the long wavelengths involved, and low frequency, have
meant this has 
been unachievable to date.

Elastic flexural waves, applicable in the mechanical
vibrations of thin elastic plates, can be
designed to have a broad zero-frequency stop-band using a periodic array of very small clamped circles. Inspired by this
experimental and theoretical 
observation, all be it in a situation far removed from seismic waves,
we 
demonstrate that it is possible to 
achieve elastic surface (Rayleigh) and body (pressure P and shear S)
wave reflectors at very large wavelengths in structured soils modelled
as a fully elastic layer periodically clamped to bedrock.
 We identify zero
frequency stop-bands that only exist in the limit of columns of
 concrete clamped at their base to the bedrock. In a realistic
configuration of a sedimentary basin 15 meters deep
we observe a zero frequency stop-band covering a broad frequency
range of $0$ to $30$ Hz. 



\end{abstract}


\maketitle

\section{Introduction}
The desire to deflect, absorb or redirect waves is ubiquitous across
many fields: electromagnetics, optics, hydrodynamics, acoustics and
elasticity. Various techniques have been developed, often in one field
but not in the others, and we are going to draw upon advances in
optics and electromagnetic wave systems to develop a methodology for
 reflecting long-wavelength seismic waves in sedimentary basins. The
 motivations to do so are clear: According to the US Geological Survey 
there are millions of earthquakes every year worldwide, the vast
majority are magnitude 3.9 or lower but more than 1000 measure 5.0 or
higher on the Richter scale \cite{reitherman12a}. Ground vibrations, caused by even minor
earthquakes, have an impact upon the structural integrity of buildings
and similarly intrusive ground vibrations from urban train systems,
subways, machinery such as piledrivers and roads often affect property
values or land usage. These vibrations are not simply a nuisance, but
small magnitude vibration due to machinery, or nearby railway lines,
can cause significant damage to buildings, especially over
time \cite{brule12a}. Furthermore, for buildings such as nuclear power plants and oil refineries, even a small level of damage can have disastrous consequences. Seismic waves consist of surface waves
(elliptically polarized Rayleigh waves and horizontally polarized Love waves), pressure bulk waves and shear  bulk waves; surface waves cause the
majority of any damage and travel farthest, but bulk pressure, and
shear, waves also cause damage, especially where wave trapping occurs
in sedimentary basins (so-called seismic site effects). Designing a defence structure to prevent seismic
waves from reaching buildings is therefore of substantial interest
particularly for long waves with frequency in the range $1$ to $10$ Hz
as this corresponds to the resonant fundamental frequency of many
man-made structures \cite{chopra12a}. It is, of course, these long,
low frequency, waves that are the hardest to develop protection
measures against and it is an open problem to develop such devices.

Since the late 1980s, in optics, researchers have taken advantage of
technological improvements in structuring matter to achieve control
over the flow of light \cite{joannopoulos08a,zolla05a}. The photonic
crystals typically used are periodic man-made
structures that inhibit emission due to 
band gaps \cite{john87a,yablonovitch87a}, i.e., ranges of frequency in which light
cannot propagate through the structure. The existence of stop-bands is well
predicted by Floquet-Bloch theory, which can be applied to any type of
waves propagating within periodic media \cite{wilcox78a,conca95a,gazalet2013}. Importantly, for
practical implementation, periodicity need not be perfect to preserve existence of stop-bands \cite{chen07a}.
For such periodic structures, striking effects such as slow light can be achieved on edges of stop-bands where the
wave group velocity vanishes \cite{figotin06a}. More recently the
field of 
 metamaterials has emerged that 
uses periodic arrangements of elements with size much smaller than the
considered wavelength (typically hundreds of nanometers) that acquire
effective properties of materials with negative optical index
\cite{pendry00a,sar_rpp05},
  or highly anisotropic materials such as hyperbolic metamaterials
\cite{iorch13a} or can be used to create invisibility cloaking devices
\cite{pendry06a,leonhardt06}. 

These ideas have
been translated to acoustics and the corresponding acoustic phononic crystals
have had success with early work by Sigalas and Economou \cite{economou93a}
showing that an infinite 2D array of high-density parallel cylinders
embedded in a low-density host material  possesses a complete band gap
in two dimensions; these effects are beautifully illustrated by the 
sound attenuation through a sculpture by artist Eusebio Sempere
\cite{meseguer95}, 
that exhibits partial stop-bands from $1.5$ to $4.5$ kHz, which has
had impact both on the scientific
community and the general public.
Aside from acoustics, these ideas have had
applications in other wave systems such as proposed breakwater
devices, Hu and Chan \cite{hu05a},  for surface ocean waves as a potential application of photonic
crystals at the meter-scale. In the same spirit, some of us
envisioned rerouting ocean waves around a region of still
water surrounded by concentric arrays of pillars \cite{farhat08b};
non-overtopping dykes for ocean waves can be also envisaged with meter
scale invisibility carpets for water waves \cite{dupont15a,berraquero13a}. 

Elastic phononic crystals can also be envisaged and typically
  fall into studies of flexural thin elastic plates, bulk media or
  surface waves atop thick elastic substrates; the latter supporting
  Rayleigh surfaces waves. There are close analogies with
  electromagnetic surface waves, as an example, surface plasmons which
  have features close to those of Rayleigh waves at least
  qualitatively. 
\cite{evans07a} showed the existence of Rayleigh-Bloch waves in linear
periodic gratings for flexural waves and these also appear in the full
elastodynamic setting \cite{colquitt15a}. These Rayleigh-Bloch waves
are waves localised to the grating, exponentially decaying away from
the grating, and only exist due to the periodicity, these have direct
parallels to spoof plasmons in the field of plasmonics, which is devoted to the control of surface electromagnetic waves
in structured metals \cite{maier10,pendry04a}. Thus qualitatively
ideas from electromagnetism transfer to the elastic cases and can act
as strong motivation. 

This activity has motivated 
experiments on the control of surface acoustic waves on the microscale in
phononic crystals, with holes, have been performed by Benchabane et
al. \cite{benchabane06a} and extended to the hypersonic regime
\cite{benchabane11a}. Protrusions such as pillars atop an elastic substrate
have also been considered \cite{khelif10a,achaoui11a} inspiring
studies using negative refraction for Rayleigh  \cite{khelif12} and
Lamb \cite{pierre10,dubois13,antonakakis14a} waves. Such interactions
between resonators on the surface, or surface defects, and surface
waves have found widespread application in phononic membranes \cite{graczykowski15a}
and in interrogating the contact adhesion of microspheres \cite{boechler13a} to name but a few. 
Interestingly, the existence of a 
zero-frequency stop-band in periodically pinned plates was proposed to
shield Lamb waves of very large wavelengths in \cite{antonakakis14a},
whereas experiments with arrays of thin elastic rods atop thin plates
showed deeply subwavelength shielding and localization effects in
\cite{colombi14a}. These lead one to consider the implications on
large-scale, i.e. meters, in terms of geophysics.  

Returning to seismic waves and ground vibration, Rayleigh wave
attenuation was achieved back in 1999 \cite{meseguer99a} in a marble quarry with air holes
displaying kHz stop-bands; this is at a frequency
range far higher than that required by the seismic application. The
theoretical concept of a seismic 2D grid of inclusions in the soil
interacting with a part of the earthquake signal was first created in
the experiments of \cite{brule14a} and from hereon we too envisage a 1D or 2D
phononic crystal as a 1D or 2D structured soil (natural or
artificial).  In 2012, with the aim of demonstrating the feasibility
of the concept with field data, full-scale seismic tests were held by
the M\'enard company in France using a grid of vertical empty
cylindrical holes with a 50 Hz source \cite{brule14a} that falls in
the partial stop-band of the large-scale phononic crystal. This is,
again, still too high for seismic applications, ideally one desires a
zero-frequency stop-band structure capable of attenuating long-waves
at very low frequencies. Interestingly, Miniaci et al. \cite{miniaci16a} propose using
cross-shaped, hollow and locally resonant (with rubber, steel and
concrete), cylinders to attenuate both Rayleigh and bulk waves in the
1 to 10 Hz frequency range. Unfortunately, the main drawback of this
type of locally resonant structure is the difficulty in obtaining very
large efficient stop bands; there is always a trade-off between the
relative bandwidth and the efficiency of the attenuation, which is
directly linked to the quality factor of the resonators. The frequency bandwidth of wave protection can be
 enlarged by considering arrays of resonant cylinders with different
 eigenfrequency for two-dimensional stop-bands \cite{krodel15a}, or
cubic arrays of resonant spheres \cite{achaoui16a} for 3D stop-bands, but such mechanical metamaterials
 would be hard to implement at the civil engineering scale. Buried
 isochronous mechanical oscillators have been also envisaged to filter
 the S waves of earthquakes \cite{finocchio14a}. 
A complementary approach employed in \cite{colombi16a,colombi16b} is
to draw upon the metamaterials literature that utilises subwavelength
resonators arranged, in their case, upon the surface of an elastic
half-space; this results in surface to bulk wave conversion, and
surface wave filters with band-gaps but again at higher frequency than those
required for seismic protection.

\begin{figure}[h!]
\includegraphics[width=8cm]{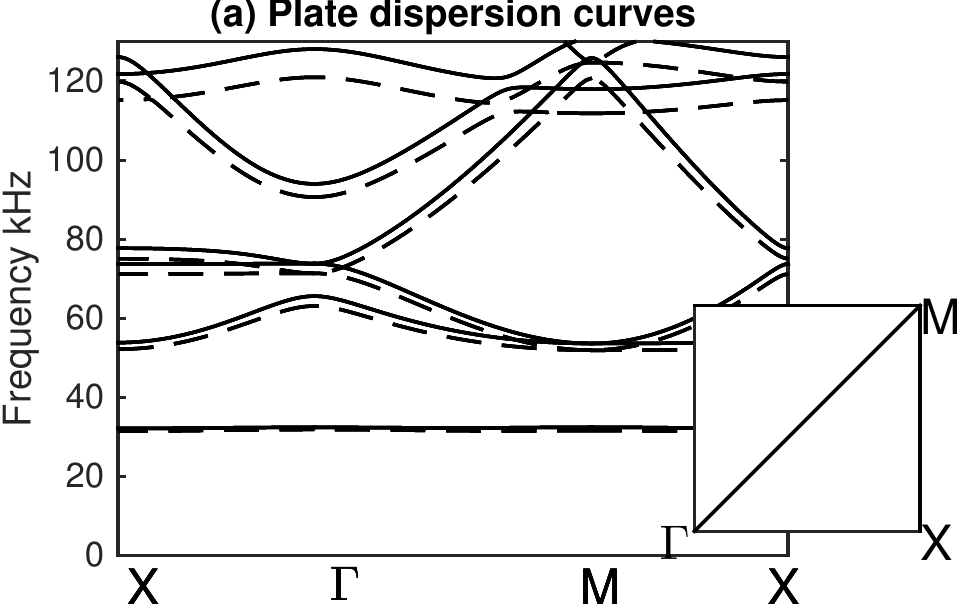}
\qquad
\addtocounter{subfigure}{1}
\subfloat[Thin plate]{\includegraphics[width=8cm]{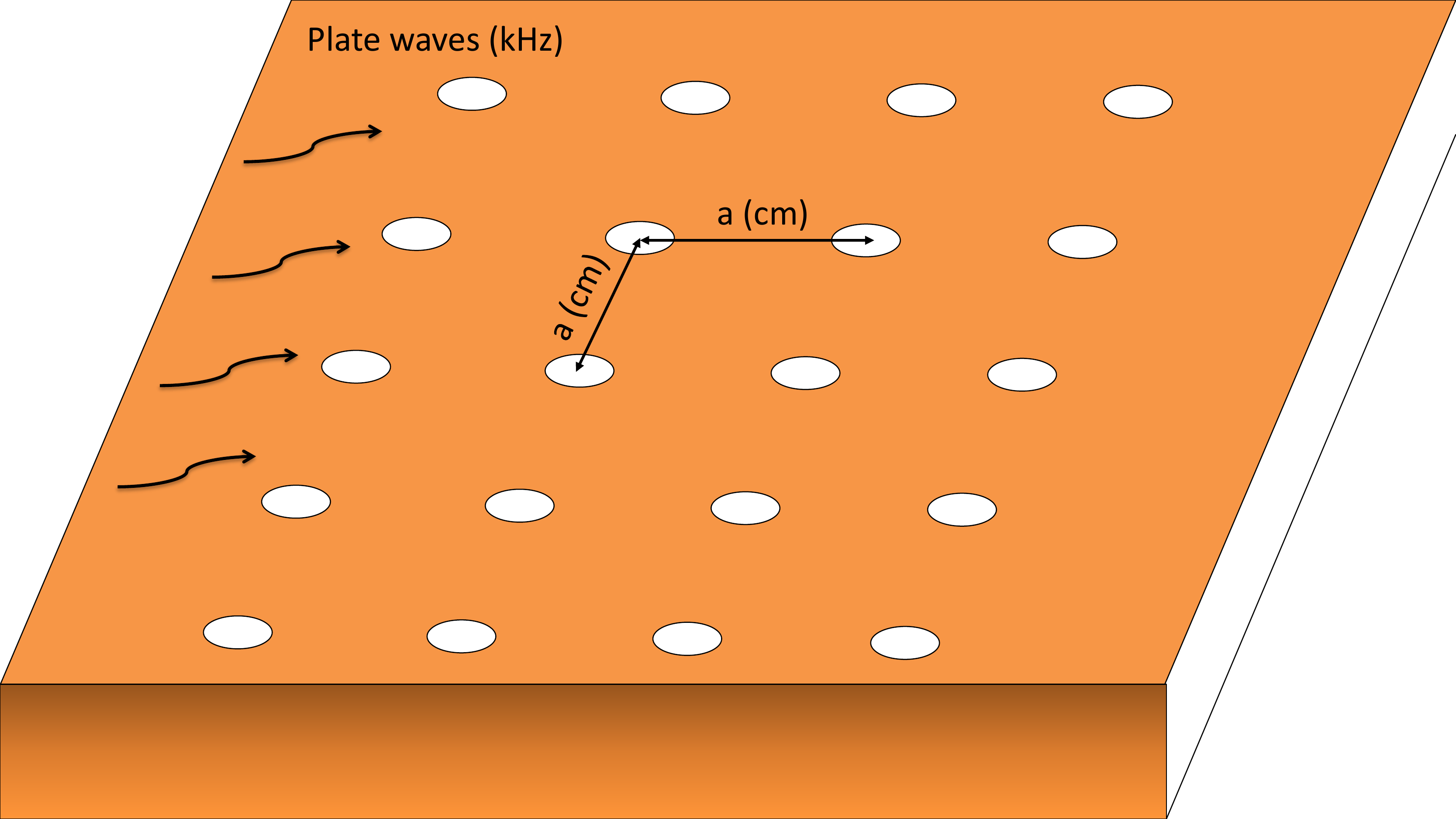}}
\caption{Clamped structured elastic plates: (a) The dispersion band
  diagram showing results from Kirchhoff-Love thin plate theory (solid) and full vector
  elastic modelling (dashed). (b) A schematic of the geometry for a
  thin clamped plate. 
} 
\label{fig:KLandElastic}
\end{figure}

Identifying that the main objective of a seismic metamaterial must be
to achieve a broad low-frequency stop-band, or even better a
zero-frequency stop-band, we turn to the apparently, distantly connected, field of thin elastic
structured plates; these are of interest in terms of flexural
waves connected with the vibration of shells. The simplest model, the
thin plate model of Kirchhoff and Love is well-known
\cite{landau70a,graff75a} and is a fourth-order scalar partial
differential equation 
for the vertical displacement; this is a
dramatic simplification over the full vector elastic system \cite{achenbach84a}
and, for
homogeneous plates, this simpler model is, in theory, valid for very thin
plates with a thickness less than $1/20^{th}$
of the typical wavelength \cite{rose04a}. None the less surprising accuracy can be
obtained in structured plates, where this thickness restriction can be relaxed, 
and recent detailed comparison of theory and experiment \cite{lefebvre17a} demonstrate that conclusions from the
Kirchhoff-Love (K-L) theory carry across into thin plate modelling even at
much higher frequencies where it might naively be thought to be
invalid (typically in practice a plate thickness less that $1/5$th of the
wavelength provides good approximations for homogeneous plates); Fig. \ref{fig:KLandElastic} shows
dispersion curves from K-L theory alongside those of full-elasticity,
both sets calculated using finite element (FE) software
\cite{comsol}. These are shown 
for a structured plate consisting of small clamped circular regions
($1.5$mm radius) arranged
on a square lattice of pitch $1$cm and for a 0.5mm thick plate of
  duraluminium ($\rho=2789$ kg/m$^3$, $E=74$ GPa, $\nu=0.33$).
 An important, indeed critical for our purposes here, observation, is
that this system has a zero-frequency stop-band. As noted earlier
there is 
close correspondence of full vector elastic calculations with those from K-L
theory far outside the range of frequencies one might usually
associate with thin plate K-L theory; the reason, as noted, in
\cite{lefebvre17a} is that what actually matters is not a constraint
from homogeneous plates, but how the wavelengths in the periodic
structured system compare to the plate thickness and they are actually 
large. Fig. \ref{fig:KLandElastic} takes advantage of the periodicity
 of the structured elastic plate, 
 as is well-known in solid state physics \cite{kittel96a}
Bloch's theorem means that, for an infinite array, one need only consider the wavenumbers in
the irreducible Brillouin zone (IBZ) which, for a square lattice, are those
in the triangle $\Gamma XM$ ($\Gamma=(0,0)$, $X=(\pi/a,0)$,
$M=(\pi/a,\pi/a)$) shown as the inset to
Fig. \ref{fig:KLandElastic}(a) and for clarity we show the frequency
dependence versus wavenumbers going around the edges of the IBZ. We
will exclusively use square lattices in this article and comment upon
other lattice geometries in section \ref{sec:discuss} (see also Fig.  \ref{fig:honeycomb}). 

\begin{figure}[t!]
\subfloat[Side view]{
\includegraphics[width=8cm]{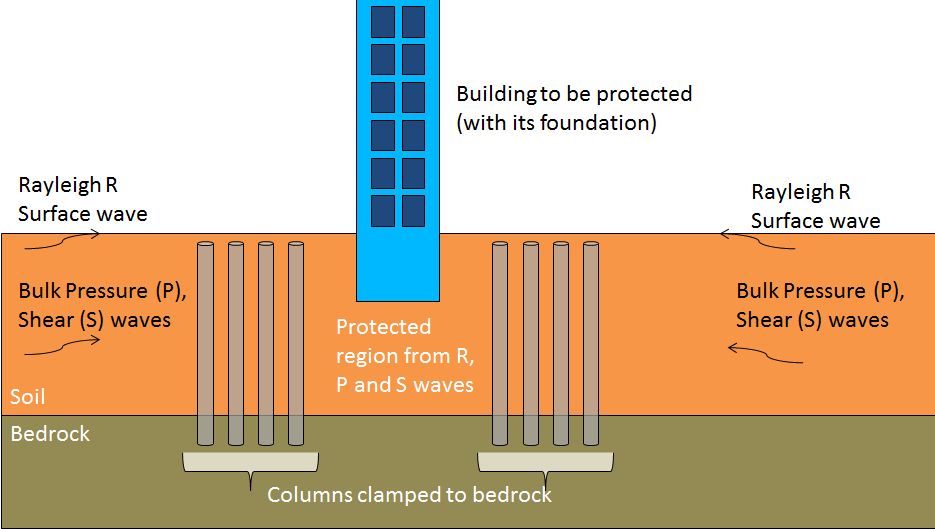}}
\qquad
\subfloat[Plan view]{\includegraphics[width=8cm]{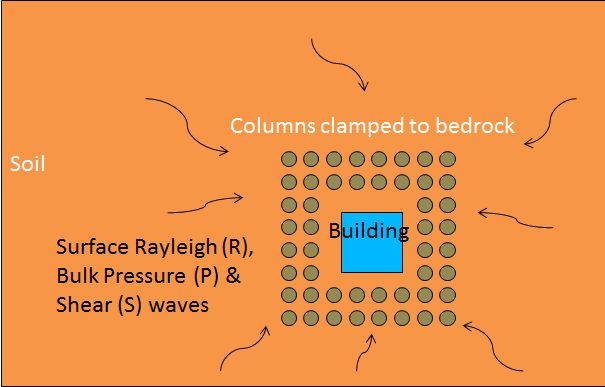}}
\caption{The seismic metamaterial consisting of columns clamped to the
bedrock surrounding the building to be protected.}
\label{fig:schematic}
\end{figure}

Having successfully identified a scenario, all be it in a different situation of
thin elastic plates far removed from the seismic application of thick
elastic substrates, giving a zero-frequency band gap we use this to
 design a seismic metamaterial to have these
characteristics \cite{colquitt16a}. The aim of this article is to investigate structuring
soil, as shown in
Fig. \ref{fig:schematic},  to protect a building or portion of the
surface.  The structuration consists of columns, in a layer of
soil, that are clamped to underlying bedrock; the columns are arranged
 in a periodic fashion. The key difference from all
previous studies is that, motivated by thin plate calculations of
Fig. \ref{fig:KLandElastic} and the resultant zero frequency band gap, we
consider the influence of clamping columns to the bedrock. 

\section{Results}

We note that many seismic applications such as ground reinforcement,
concern layers of softer soils 
overlying more solid bedrock and modern civil engineering processes
allow columns to be clamped, that is, rigidly attached, to the
bedrock \cite{brule17a}. The layers are no longer thin enough to employ K-L
thin plate theory and the effect of depth (equivalently thickness of
the plate) is now important. We explore the potential of seismic
metamaterial devices shown in Fig. \ref{fig:schematic}; the side view shows a structure
atop a soil layer of finite thickness that overlays the bedrock;
columns clamped to the bedrock puncture the soil layer and reach to
the surface, or close to surface, a top view shows the array of
columns encircling the building to be protected.

We consider small strain seismicity, \cite{semblat09}, with a rate
of deformation ($\gamma$ such that $\gamma\ll 10^{-4}$), in this situation the duration of seismic
disturbance, for this type of earthquake, is sufficiently
short to accept the hypothesis of elastic behaviour for the soil. 
 We can therefore take realistic soil, rock and depth parameters from geophysics with
the Young's modulus, $E=3K(1-2\nu)$, of the soil and rock being $153$ MPa and $30$
GPa respectively, Poisson ratios of $\nu=0.3$ for both and density $\rho$
of the soil and rock being $1800$ kg/m$^3$ and $2500$ kg/m$^3$
respectively. 
A typical depth of soil is $15$~m, which 
overlays bedrock of depth $5$~m with effectively rigid material
beneath it  (i.e. the bedrock interface is modelled with Dirichlet
boundary data, which is zero 
 displacement there). 

Given the geometric and physical complexity we proceed to numerical
simulations. 
For these simulations we solve the elastic Navier equation \cite{ciarlet97} with time-harmonic
motion considered at fixed frequency i.e. in the absence of any sources we consider
\begin{eqnarray}
\nabla\cdot \left[  {\bf C}({\bf x}) :\nabla {\bf  u}({\bf x})
 \right]  + \rho({\bf x})\omega^2 {\bf  u}({\bf x})={\bf 0} \; ,
\label{navier}
\end{eqnarray}
where ${\bf C}$ is an isotropic symmetric rank-4 elasticity tensor, with spatially varying entries written in a Cartesian basis
$C_{ijkl}=K\delta_{ij}\delta_{kl}+\mu\left(\delta_{ik}\delta_{jl}+\delta_{il}\delta_{jk}-2/3\delta_{ij}\delta_{kl}\right)$,
with $\delta_{ij}$ the Kronecker symbol and $K$ the bulk modulus, $\mu$ the shear modulus, $\rho$ the density, $\omega$
the angular wave frequency and ${\bf  u}({\bf x})=(u_1,u_2,u_3)(x_1,x_2,x_3)$ the 3 component displacement field.
We discretize the weak form of (\ref{navier}) using finite
element methods and in particular we utilise COMSOL
\cite{comsol}.
For  band structure calculations, we take advantage of the periodicity of
 the system in the horizontal $(x_1,x_2)$-plane to consider a single elementary cell. For this we use the
 Bloch-Floquet theorem i.e. we assume that solutions of (\ref{navier}) are such that
\begin{eqnarray}
{\bf  u}({\bf x}+{\bf a})={\bf  u}({\bf x})\exp(i {\bf k}\cdot{\bf a})
\label{bloch}
\end{eqnarray}
 where ${\bf a}=(a_1,a_2,0)$ is the lattice vector of the 2D array of columns and ${\bf k}=(k_1,k_2,0)$ is the Bloch (or momentum) wavevector in 2D reciprocal space (both ${\bf a}$ and ${\bf k}$ lie in the horizontal plane).

We then construct, and solve, the eigenvalue problem created by
substituting the Bloch wave form ${\bf u}({\bf x})={\bf U}_{\bf
  k}({\bf x})\exp(i{\bf k}\cdot{\bf x})$ in (\ref{navier}); this leads to the eigenvalue problem for the shifted Navier operator ${\cal N}_{\bf k}$:
\begin{equation}
{\cal N}_{\bf k}({\bf  U}_{\bf k}({\bf x}))=\rho({\bf x}){(\omega^{\bf k})}^2{\bf  U}_{\bf k}({\bf x})
\label{navierbloch}
\end{equation}
where ${\cal N}_{\bf k}=-\{\nabla+i{\bf k}\}\cdot [  {\bf C}({\bf x}) : \{ \nabla +i{\bf k}\} ]$ parameterized by ${\bf k}$, which relates the phase shift across the cell to the
 frequency $\omega^{\bf k}$  (see \cite{nicolet04a} for the electromagnetic case). For a given ${\bf k}$, one gets a discrete set of eigenfrequencies $0\leq\omega_1^{\bf k}\leq \omega_2^{\bf k}\leq \cdots \omega_n^{\bf k} \cdots$ tending to infinity as the operator ${\cal N}_{\bf k}$ in (\ref{navierbloch}) has a compact resolvent. Since the eigenfrequencies $\omega^{\bf k}$ are continuous with respect to ${\bf k}$, when ${\bf k}$ varies in the Brillouin zone $BZ=[-\pi/a_1,\pi/a_1]\times[-\pi/a_2,\pi/a_2]$, the so-called Bloch band spectrum is found
\begin{equation}
\sigma_{Bloch}=\{0\}\cup\bigcup_{j\geq 1}[\min_{{\bf k} \in BZ}\omega_j^{\bf k},\max_{{\bf k} \in BZ}\omega_j^{\bf k}]
\end{equation}
see for instance \cite{conca95a,zolla05a}. The dispersion surfaces computed are critical in guiding our understanding of the Bloch wave behaviour, we refer to \cite{lefebvre17a} for structured media displaying elliptic, parabolic and
hyperbolic shapes of dispersion surfaces leading to extreme elastic wave control at certain frequencies revealed by so-called High-Frequency Homogenization \cite{craster10a}. Importantly,
when $\mid{\bf k}\mid\to 0$, the quasi-static limit of the band
spectrum allows one to deduce effective properties of the periodic
structure (such as anisotropy) through the slope or curvature of
dispersion surfaces in the neighbourhood of $\Gamma$ point
\cite{bensoussan78a,conca95a} depending upon whether they have a
conical or parabolic shape. One can then interpret the effective behaviour of the low frequency Bloch waves in terms of their effective group velocity, or effective mass \cite{guenneau03a}. However, when some Dirichlet data, ${\bf U}_{\bf k}={\bf 0}$, is set on some domain within the IBZ, corresponding to a clamped (possibly very small) inclusion, one infers from the maximum principle \cite{bensoussan78a,conca95a} that ${\bf U}_{\bf k}$ is null everywhere in the IBZ, and is not an admissible eigenfield. Hence $\{0\}\not\subset\sigma_{Bloch}$. This seemingly minor remark has important practical consequences: it shows that a periodically constrained elastic structure has a zero-frequency stop-band, which can be used to reflect low frequency elastic waves. 

Making use of the symmetries of the periodic structure, it is common
to consider the dispersion properties of Bloch waves by letting ${\bf
  k}$ vary only along the edges of the irreducible Brillouin zone IBZ
\cite{joannopoulos08a}, and so one need only compute dispersion
curves, and no longer dispersion surfaces, which greatly reduces
computational effort. However, this needs to be done carefully,
otherwise edges of stop-bands might be erroneously estimated
\cite{harrison07a,craster12a,brule16a}. The computational effort can
be further reduced in the case of periodic thin-plates when one looks
for eigenvalues and eigenfunctions of the so-called Kirchhoff-Love
operator \cite{ciarlet97} ${\cal L}_{\bf k}=-{[\{\nabla_H+i{\bf k}\}\cdot \{ \nabla_H
  +i{\bf k}\}]}^2$ 
\begin{equation}
{\cal L}_{\bf k}(U_3^{\bf k})=\frac{12(1-\nu^2)\rho({\bf x})}{Eh^2}{(\omega^{\bf k})^2}{U_3}^{\bf k}({\bf x})
\label{lovebloch}
\end{equation}
where $h$ is the plate thickness. An important advantage of the K-L
equations is that the $x_3$ variable is no longer present, the
equations are scalar and one only needs to consider the $x_1, x_2$
coordinates and the horizontal Laplacian
$\nabla_H=(\partial/\partial x_1,\partial/\partial x_2,0)$. 
 From the eigenvalue problem (\ref{lovebloch}), one can infer the
 dispersion curves associated with Bloch flexural waves associated
 with vertical displacements $(0,0,U^{\bf k}_3)$ in homogeneous
 thin-plates with clamped inclusions; the corresponding solid
 dispersion curves in Fig. \ref{fig:KLandElastic} almost match the
 dashed ones computed from (\ref{navierbloch}); the discrepancy
 between the dispersion curves computed from K-L (\ref{lovebloch}) and
 full elasticity  (\ref{navierbloch}) operators does increase with
 frequency but not to the extent one would expect from the homogeneous
 plate theory. Fig. \ref{fig:KLandElastic} uses typical parameters, that is, an infinite periodic square array (i.e. $a_1=a_2=a$) taken to have $1$~cm pitch $\mid{\bf a}\mid=a$  for a $0.5$mm thick plate of duraluminium ($\rho=2789$ kg/m$^3$, $E=74$ GPa, $\nu=0.33$). Note that the zero frequency stop-band in Fig. \ref{fig:KLandElastic}(b) ranges from $0$ to above $30$ KHz, although the inclusions are fairly small ($1.5$mm radius). If we now consider that the array pitch is one thousand times larger i.e. $a=10$m, and the plate is $0.5$m thick, with clamped inclusions $1.5$m in radius, we achieve a stop-band from 0 to above $30$ Hz. These are frequencies of interest in civil engineering, hence one might be tempted to say that a large scale periodically constrained Duraluminium plate lying atop softer soil could serve as large scale infrastructure's foundations as it would filter surface waves propagating therein. However, we shall not pursue this route here, but turn to other structured soil designs. 
 
\begin{figure}[h!]
\includegraphics[width=12cm]{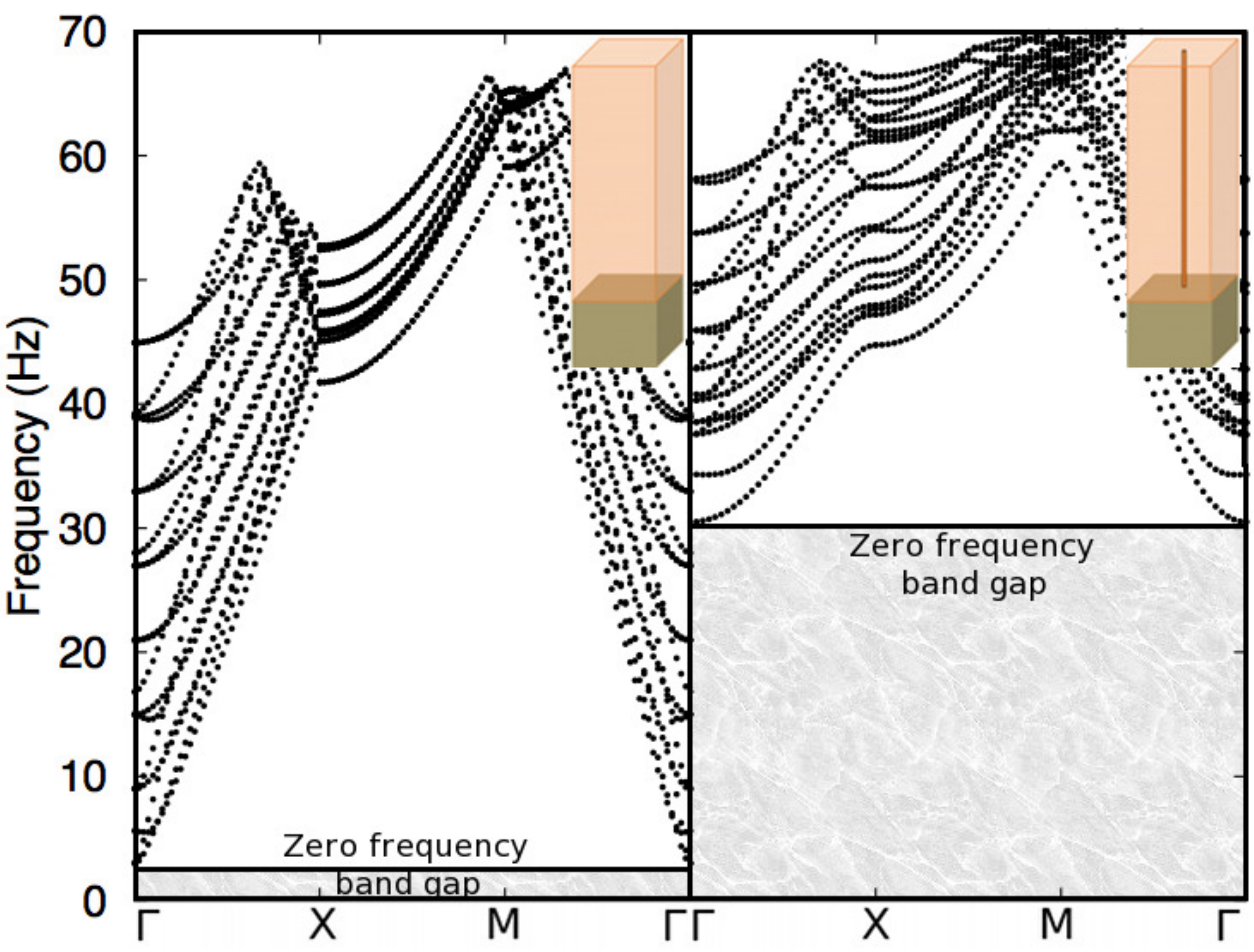}
\caption{Dispersion curves for idealised cases: (a) clamped bedrock
  with no columns
(b) a clamped perfectly rigid column (i.e. with zero displacement
assumed on all its boundary) of radius 0.15~m. 
These are shown around the edges
of the irreducible Brillouin zone (see Fig. \ref{fig:KLandElastic}) $\Gamma XM$.}
\label{fig:idealised}
\end{figure}

 First, simply taking into
account just the finite soil layer atop the bedrock, and considering waves in
this layer, leads as shown in Fig. \ref{fig:idealised}(a), to a small
zero frequency stop-band with the cut-off frequency at $2.7$~Hz. This
is to be expected: If one considers this
 layered system to be a waveguide then fixing one wall to be
rigid will automatically shift the modal cut-off frequencies and no modes will
propagate for extremely low frequencies. 
One point of note is that
results, not shown, also demonstrate a weak dependence upon soil depth
with the cut-off frequency rising as the depth decreases and vice-versa. 
Although instructive, 
this case does not lead to the desired wide zero-frequency stop-band.  Turning now to
another idealised situation, that of pillars (of radius $0.3$~m) that are completely rigid,
and immovable all along their length, and that are buried in the bedrock, we see in
Fig. \ref{fig:idealised}(b) that an extremely wide zero-frequency stop-band
stretching all the way to 30 Hz is achieved. This is unphysical as it
is not possible to clamp a column all along its length, but it does
demonstrate that there is potential to create a very wide band gap. The other
issue is that dispersion curves are reliant upon an infinite array
and for realistic systems the array will be finite. 

\begin{figure}[h!]
\begin{center}
\includegraphics[width=14cm]{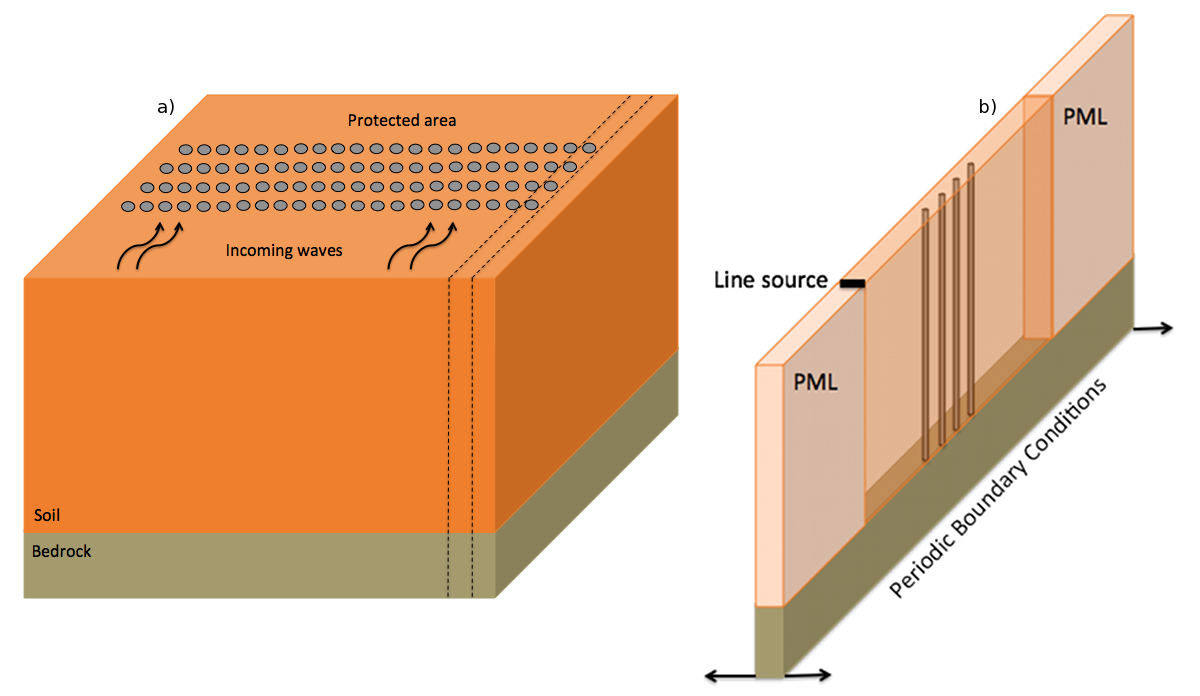}
\end{center}
\caption{The computational domain for transmission spectra. (a) Waves
  incident upon a finite array (b) for normal incidence a reduced
  domain, shown by the dashed lines in (a), can be used.}
\label{fig:domain}
\end{figure}

\begin{figure}[h!]
\includegraphics[width=15cm]{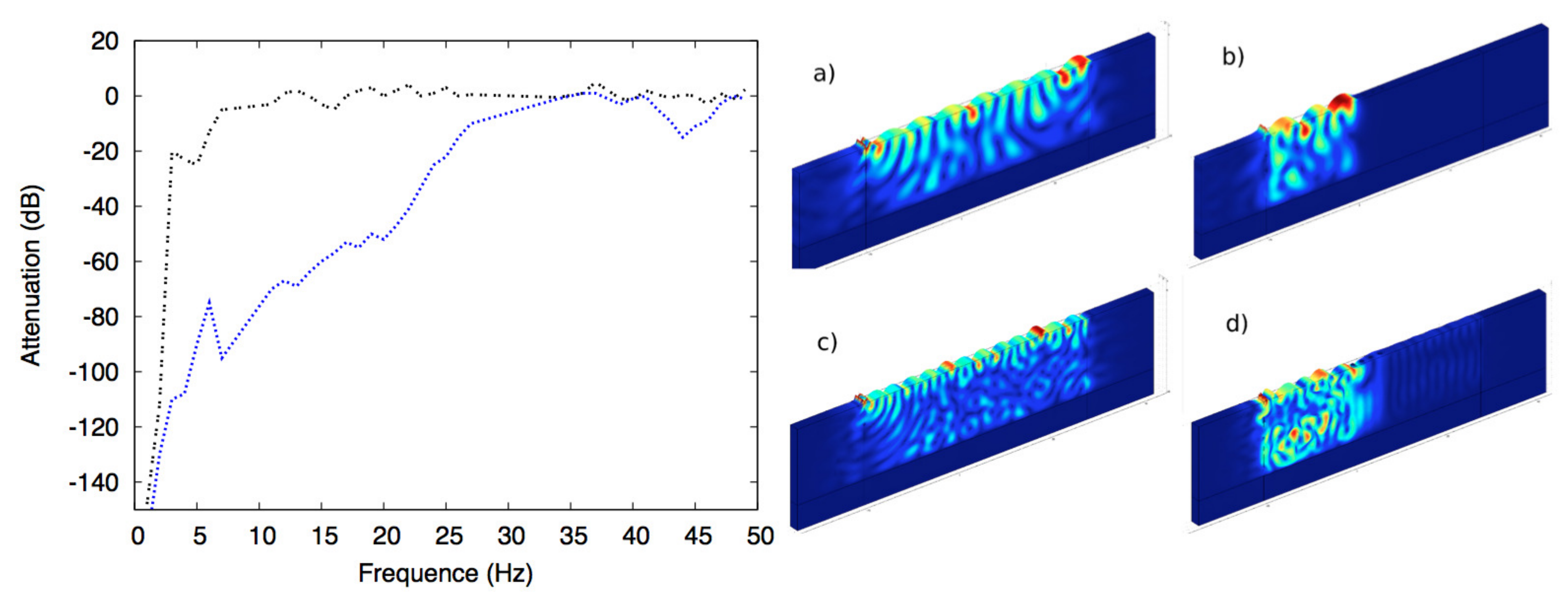}
\caption{Idealised cases: simulations and attenuation. Panels (a,c)
  show the absolute value of the vertical displacement field for the clamped bedrock
  respectively, (b,d) 
show the field for the structured soil consisting of  $4$
immovable rigid columns, of radius $0.15$~m, both simulations are  shown for at $20$ (a,b)
and $35$ Hz (c,d). The left panel shows the attenuation versus frequency
with the upper curve (dashed) for soil layer with clamped bedrock and the lower
curve (dotted) for the immovable rigid columns.}
\label{fig:ideal_trans}
\end{figure}

To further motivate the importance of a zero frequency stop-band and
illustrate its effect, we simulate normal incidence upon a finite
array protecting a region and generate the transmission spectra. The 
 computational domain is shown in
Fig. \ref{fig:domain} which consists of the clamped columns (piles) in
the soil 
attached to bedrock with waves incident from the line source. To ensure that
there is no effect from the size of the computational domain elastic perfectly
matched layers (PMLs) \cite{diatta16a} are employed to prevent spurious reflections in the
propagation and reflection directions. Since normal incidence is
studied we will consider a single row of columns
(Fig. \ref{fig:domain}(b)) and then use periodic conditions 
 spanwise. We again proceed numerically in the time harmonic situation
 and solve the elastic Navier equation using finite
element methods and in particular utilise COMSOL
\cite{comsol}  and use a line source at the surface (invariant across
the domain) to initiate either Rayleigh surface waves or SH polarized
waves; we use Rayleigh excitation in Fig. \ref{fig:ideal_trans}.

\captionsetup[subfigure]{labelformat=empty}

The effect of the clamped columns is most unequivocally seen by
looking at the transmission spectra, that is the transmission computed
at the soil/PML computational interface at the opposite side of the
columns from the line source normalised by the amplitude of
the source. The left
hand panel of 
Fig. \ref{fig:ideal_trans}  shows very clearly  that the fields are
 highly attenuated beneath the cut-off frequencies of each case, at $2.7$ and $30$ Hz
 respectively. The vertical scale is in decibels and so 
 this represents the attenuation that would be seen, and in the
 clamped column case this is for
 just four clamped columns and is highly effective at blocking
 incoming waves. This is further exemplified by examining the physical
 fields as shown in Fig. \ref{fig:ideal_trans} (a,c) for the
 clamped bedrock and (b,d) for the idealised clamped, perfectly
 immovable, columns. The effect of the columns is clearly seen in the
 fields with almost perfect reflection at the frequencies
 illustrated. 

The results above are not 
physically realisable as no columns can be completely fixed and rigid
all along
their length (i.e. zero displacement on all of its boundaries); this 
provides an ideal upper bound on the possible efficiency of this design, but
it is highly encouraging that clamped columnar structures could, at
least, achieve part of this behaviour.

\begin{figure}[h!]
\includegraphics[width=12cm]{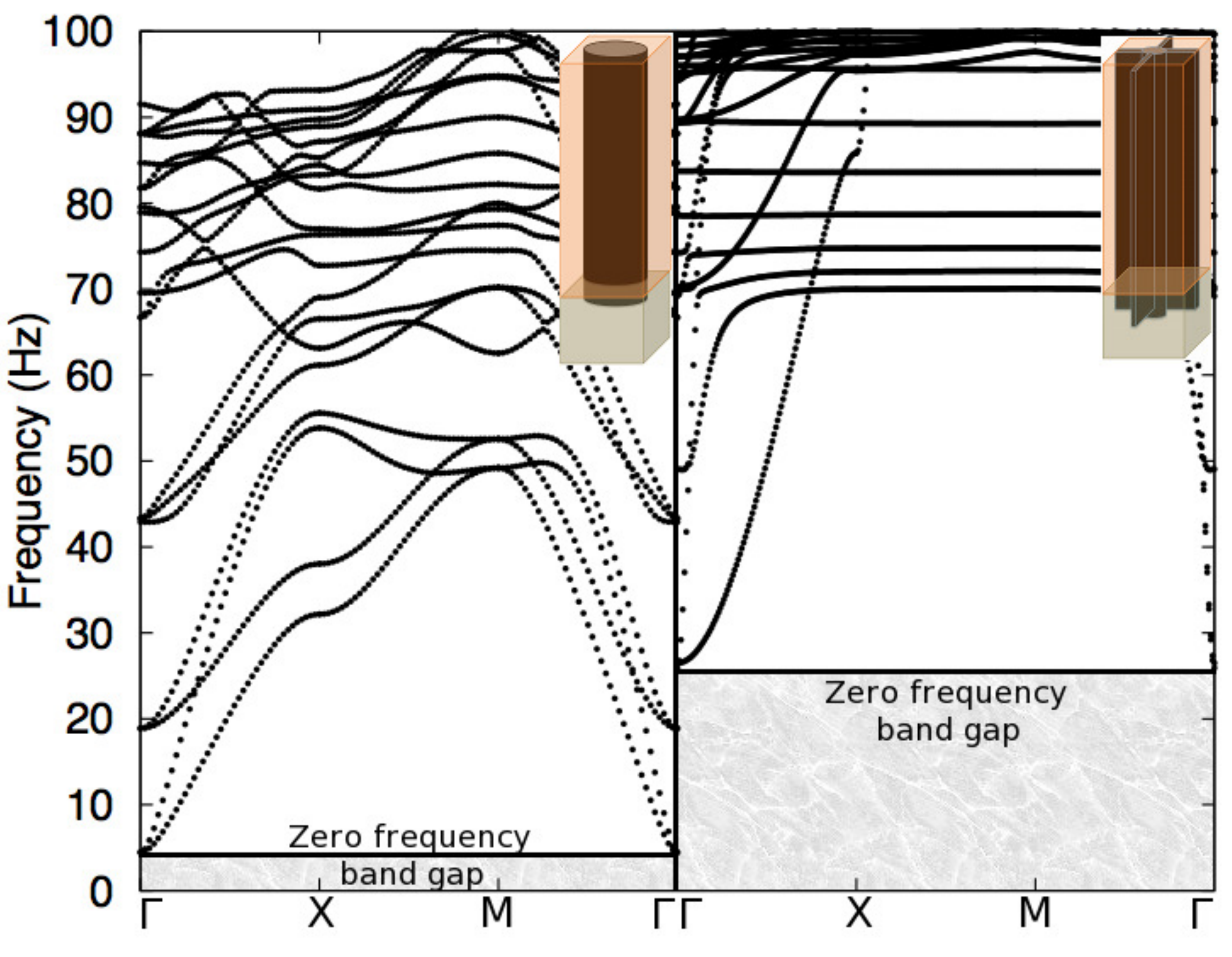}
\caption{Dispersion curves for realistic cases: (a) steel cylindrical
  columns of radius $0.6$~m 
  clamped to the bedrock 
(b) steel columns of radius $0.3$~m linked to their nearest neighbours via steel plates.  
These are shown around the edges
of the irreducible Brillouin zone $\Gamma XM$(see Fig. \ref{fig:KLandElastic}).}
\label{fig:realistic}
\end{figure}

We now turn to realistic designs using illustrative columnar materials
and geometries. We choose two examples and treat them in detail, first
cylindrical columns of steel (density 7850~kg/m$^3$, Poisson ratio
0.33), $1.2$~m in diameter, in a soil layer of $15$~m and
bedrock of depth $5$~m; the columns are clamped to the bedrock but
otherwise free to vibrate and interact with the soil and bedrock. 
For an infinite array placed at the vertices of a square array (of
pitch $2$~m) the dispersion curves are shown in
Fig. \ref{fig:realistic}(a). It is clear that this has a
zero-frequency stop-band up to $4.5$~Hz, which is not dramatic in
extent 
but is none the less an improvement over the unstructured soil. It is, however, far
from the optimal case shown in Fig. \ref{fig:idealised}.

\begin{figure}[h!]
\includegraphics[width=14cm]
{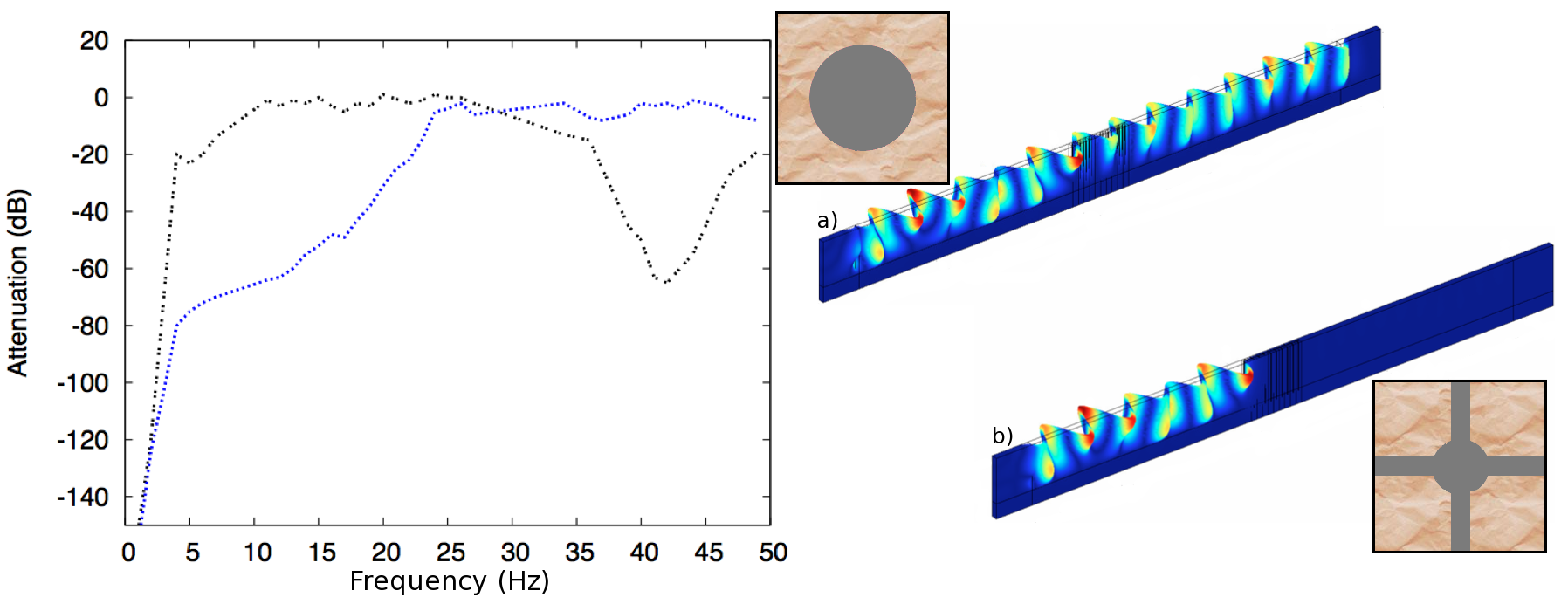}
\caption{Attenuation (in dB) of incoming shear waves versus frequency (in Hz)
through an array as in Fig. \ref{fig:domain} with 4 rows of clamped
cylindrical inclusions of radius 0.6~m
(upper curve) and clamped cross-shaped inclusions (lower curve). The
absolute value of the vertical displacement at 13Hz for (a) the steel
column and (b) a column augmented by attachments to its neighbours;
the geometries, which share the same filling fraction, are shown as
insets.}
\label{fig:fig3}
\end{figure}

\begin{figure}[h!]
\includegraphics[width=15cm]
{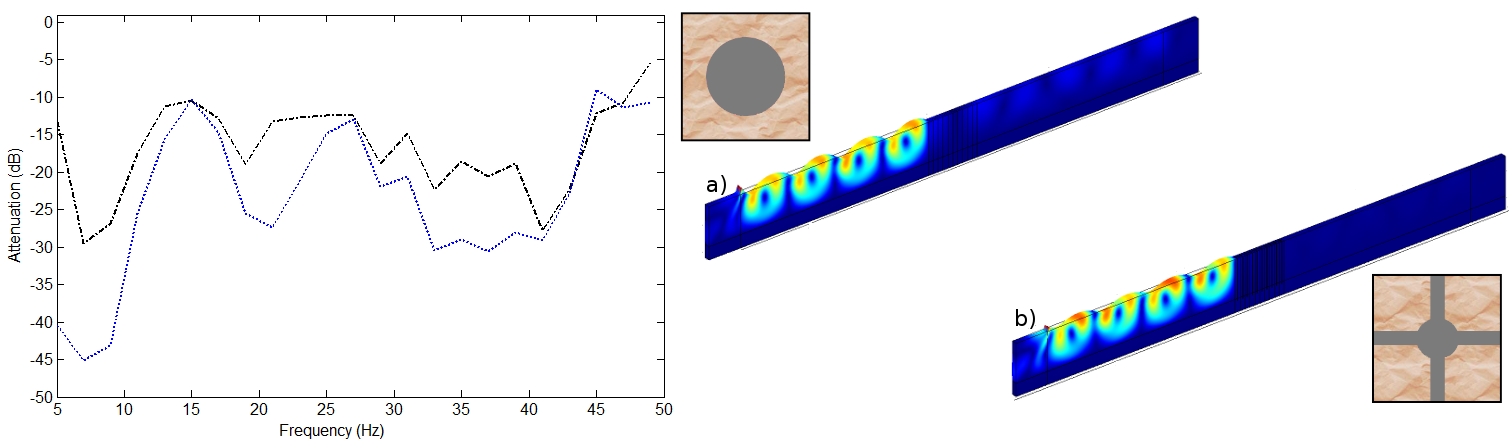}
\caption{Attenuation (in dB) of incoming Rayleigh waves versus frequency (in Hz)
through an array as in Fig. \ref{fig:domain} with 10 rows of clamped
cylindrical inclusions of radius $0.6$~m
(upper curve) and clamped cross-shaped inclusions (lower curve). The
absolute value of the vertical displacement at 9~Hz for (a) the steel
column and (b) a column augmented by attachments to its neighbours;
the geometries, which share the same filling fraction, are shown as
insets.}
\label{fig:fig3b}
\end{figure}


It is well known in the 
phononic crystal community that very broad band gaps can be designed
by the use of cross-shaped inclusions \cite{wang11a}, reminiscent of lattice models \cite{martinsson03a}, so we utilise
this shape here and further augment this by linking the columns. We consider a cross-shaped inclusion having the same
cross-sectional area as the cylinder thereby comparing inclusions
that have the same filling
fraction. In this second configuration we take a cylindrical column of radius
$0.2$~m  and  
each column is then 
linked to its nearest neighbours by struts (long steel plates of
$0.2$~m in thickness), see Fig. 
\ref{fig:realistic}(b), and this also gives the columns a structural link
to its neighbours. This additional reinforcement then leads to a
dramatic enhancement in the protected frequency range which is now up
to $26$~Hz. The dispersion curves of Fig. \ref{fig:realistic} are for
an infinite array and do not reveal the potential efficiency of the
design, for this we need the transmission spectra.

We consider both shear wave and Rayleigh wave incidence, as they have
different character. 
Shear wave incidence is illustrated in  Fig. \ref{fig:fig3} that shows the attenuation of the clamped cylinders and
the clamped cross-shaped inclusions; the latter clearly perform much
better and the attenuation is strongly enhanced below $26$Hz. Notably
this performance is for normal incidence upon a finite array only four
columns deep, and improves further if the number of columns is
increased. Movies constructed using the time harmonic solutions to
show the time evolution of an incoming wave incident upon the
columns are shown in the supplementary material. As an aside in Fig. \ref{fig:fig3} there is a dip in the transmission spectra for the clamped
cylindrical columns at approximately 40~Hz which is due to the
polarisation of the source and its interaction with the phononic
crystal; this can be identified in Fig. \ref{fig:cases}A when taking into
account the polarization state of shear waves. 

Rayleigh wave incidence is illustrated in  Fig. \ref{fig:fig3b} showing
again attenuation of the clamped cylinders and
the clamped cross-shaped inclusions and the associated displacement
fields. The attenuation is not as dramatic as for shear waves due to
the stronger
coupling of Rayleigh waves into the structure and  
the finite structure as a whole can vibrate, i.e. the signal at 15~Hz,
none the less strong attenuation is achieved for low frequencies. 

\begin{figure}[h!]
\includegraphics[width=9cm]{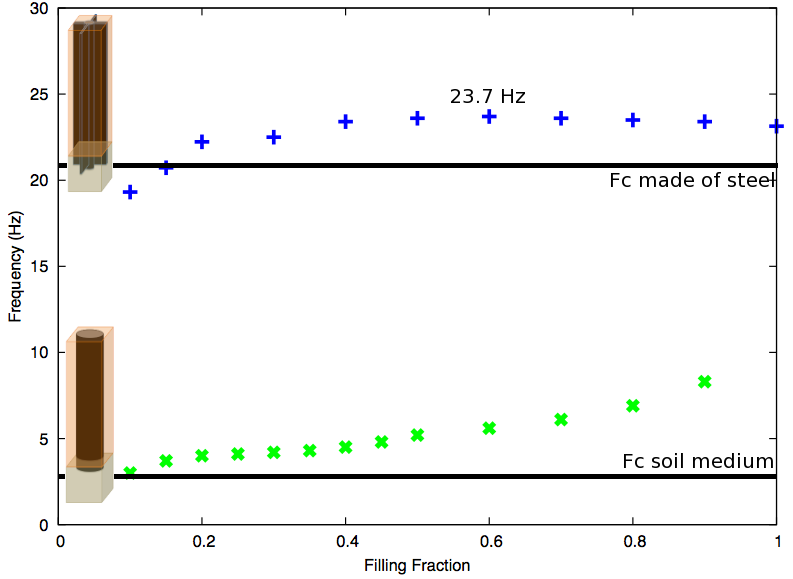}
\caption{The upper extent of the zero frequency band-gap versus
  filling fraction (i.e. soil substitution rate) for the clamped cylindrical and cross-like
  inclusions. Note the upper curve is non-monotonic.}
\label{fig:gapmap}
\end{figure}

We now address the issue of whether there is an optimal filling
fraction. In Fig. \ref{fig:gapmap} we show the upper extent of the
zero frequency band-gap for the cross-like inclusions and cylindrical
inclusions. In the latter case the upper extent increases
monotonically as the filling fraction (also called the soil
  substitution in the civil engineering literature) increases, as one
would expect. 
For the cross-like inclusion there is a non-monotonic behaviour, all
be it very gradual, with a maximum for a filling
faction of around $0.6$. In practical terms, a soil substitution rate beyond $10$ to $20$ percent is quite substantial
in civil engineering due to cost concerns, and in practice the smaller filling fractions
are more relevant.

\begin{figure}[h!]
\includegraphics[width=9cm]{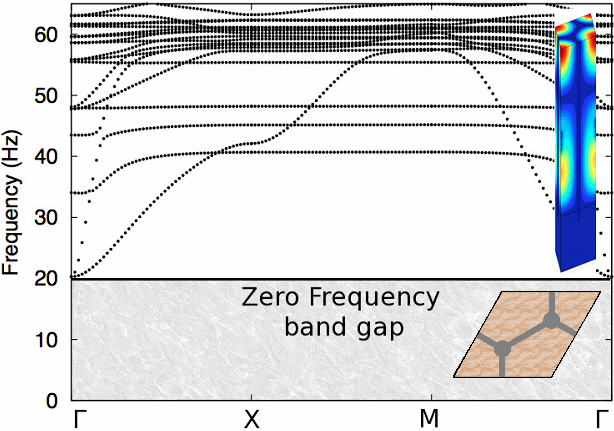}
\caption{Dispersion curves for steel columns of radius 0.3~m arranged on a honeycomb
  lattice of pitch $2\sqrt 3$ to maintain 2~m spacing between adjacent
  columns and attached by steel plates along the honeycomb
  lattice. }
\label{fig:honeycomb}
\end{figure}

It is natural, of course, given that we have presented in detail a
design on a square lattice to consider whether changing the underlying lattice
structure to another Bravais lattice creates fundamental changes. It
does not, it alters the detail of the extent of the zero frequency
band gap but not the physics behind this. As a further illustration we
show, in Fig. \ref{fig:honeycomb}, the dispersion curves for the steel
columns (radius 0.3~m and buried 80~cm in the bedrock) placed on a honeycomb lattice with plate
attachments along the honeycomb. Again one notes an extensive zero
frequency bandgap extending up to 20~Hz, thus there is no specific
advantage of using a honeycomb, instead of square, array.

\begin{figure}[h!]
\includegraphics[width=12cm]{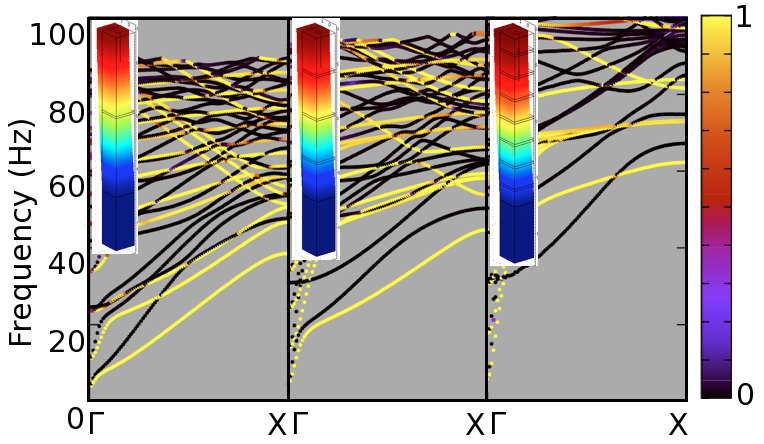}
\caption{The dispersion curves along $\Gamma X$ for a vertical steel
  column of radius 0.3~m 
 piercing horizontal steel plates of thickness
  0.2~m. From left to right we
  start with a single plate at depth $7.5$~m splitting the soil layer
  in half with then $3$ and $7$ equally spaced plates in the other two
panels. The colours of the bands show the amount of polarization,
  $p_2$, (\ref{eq:polarisation}), 
  along the propagation direction ($x_2$ axis).}
\label{fig:horizontal}
\end{figure}

\begin{figure}[h!]
\includegraphics[width=13cm]{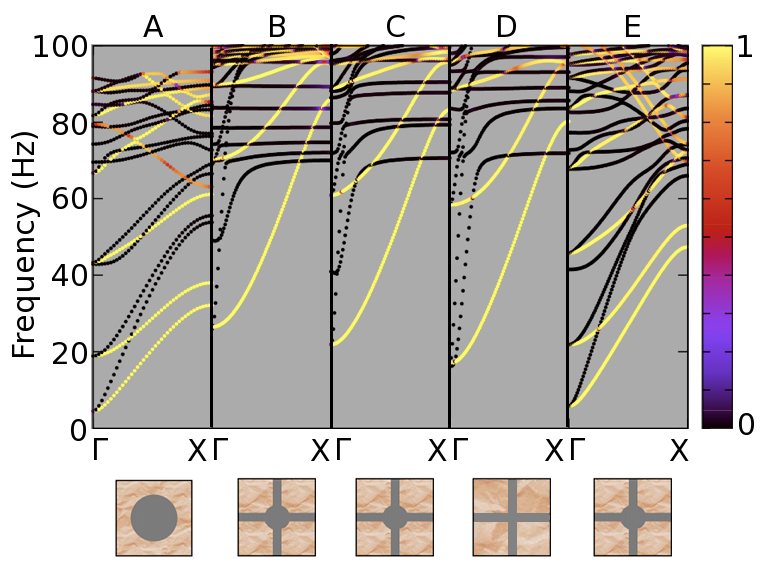}
\caption{Dispersion curves in $X\Gamma$ for 5 cases. A, is the steel column
of Fig.  6(a) buried $80$~cm in bedrock, B, is the steel column joined to its neighbours in a
square lattice, buried $80$~cm of Fig 6 (b). Panels C and D are not buried in the rock and
merely are joined at the soil / bedrock interface to the bedrock and
then have the central steel column absent respectively. E shows the
effect of 
decoupling the joining plates at the edges of the periodic cell. For
clarity the cross-section of the cell is shown beneath each case. The colours of the bands show the amount of polarization,
  $p_2$, (\ref{eq:polarisation}), 
  along the propagation direction ($x_2$ axis).}
\label{fig:cases}
\end{figure}

A key point of the design, further emphasised in \cite{achaoui2016_patent}, is
the burying of columns in the bedrock and connectivity between columns  and to demonstrate how important this is
we consider an array of horizontal steel plates of thickness 0.2~m 
 that splits the soil layer into a laminate structure and a column of
 radius 0.3~m; 
the band diagrams for this structure are shown in
Fig. \ref{fig:horizontal} that also show the polarization $p_2$ (along $x_2$)
defined as 
\beq
 p_2^2=\frac{1}{V}\int_{-a_1}^{a_1}\int_{-a_2}^{a_2}\int_0^h\frac{\vert u_2\vert^2}{\vert u_1\vert^2 +\vert u_2\vert^2
   +\vert u_3\vert^2} d{x_1}d{x_2}d{x_3}
\label{eq:polarisation}
\eeq
where the triple integral is over the periodic cell $[-a_1,a_1]\times [-a_2,a_2]\times[0,h]$ with $a_1=a_2=1$ m the array pitch, $h=15$ m the soil depth and $V$ the volume of the cell (note that we checked the displacement field vanishes in the bedrock). Here there is no zero frequency band gap
with the dispersion curves always passing through the origin, however
there are some interesting features observable in this diagram. For
instance, there are pronounced changes of curvature in the lower bands
 that may be of interest and as one increases the lamination the
regions of wavenumber space for which there is propagation shrink. 

Finally, we investigate the sensitivity of the upper-extent of the
zero-frequency band gap to attachment in the bedrock and to modifying
the square lattice structure of a steel rod attached to its neighbours
by stell plates. We do this in Fig. \ref{fig:cases} and show the
dispersion cuves only in the $\Gamma X$, that is for normal
propagation to an array, and consider 5 cases: A, is the steel column
of radius $0.6$~m 
of Fig. \ref{fig:realistic}(a) buried $80$~cm in bedrock showing a small zero-frequency band-gap, B, is the
reference case of the steel column of radius $0.3$~m joined to its neighbours in a
square lattice, buried $80$~cm of Fig. \ref{fig:realistic}(b) and sharing the same
filling fraction as A. Panels C and D are not buried in the rock and
merely are joined at the soil / bedrock interface to the bedrock and
then have the central steel column absent respectively; the upper
extent of the band-gap decreases. E goes one stage further and
decouples the joining plates at the edges of the periodic cell and
then the band-gap width decreases dramatically. It is clear that the
combination of having the steel column, buried in the bedrock, and the
coupling of each column to its neighbours via the steel plates is
essential to obtain the broad ultra-low frequency stop-bands.

\section{Discussion}
\label{sec:discuss}

In Geotechnics, so-called composite soils made of inclusions inserted
in a matrix of soil are mainly developed using a pseudo-static
analysis; the dynamic loading is converted into an equivalent static
loading and the objective is to obtain a higher value of Young’s
modulus or shear modulus G (or $\mu$) in order to reduce the
deformation of the soil-deep foundation system
\cite{afps12,pecker09}. Homogenization techniques have been also
applied to composite soils \cite{debuhan86,gueguin14} and recently
developed for dynamic loading \cite{nguyen14}. The emerging field of
seismic metamaterials, based on wave physics, enables us to revisit
several longstanding problems of earthquake protection from this
fully dynamic point of view.

In this vein, we have considered here how one could use these physics-based
ideas to protect specific areas from low-frequency vibration. We have demonstrated, conclusively, that it is possible to design
realistic seismic metamaterial devices, in the sense of the structures
that form the material being subwavelength and in the limit  of small
strains such that we have linear elastic conditions.
 The desired performance is achieved by
creating zero-frequency stop-bands using clamping of columns to
underlying bedrock. Clearly the challenge is now to build such
structures in the spirit of \cite{brule14a} and we anticipate that
these results will motivate large-scale experiments to verify this
since there are potential applications in seismic defence structures \cite{achaoui2016_patent}. 
It is notable that zero-frequency stop-bands will
arise in other areas of physics whenever periodic arrays of 
inclusions have zero field, i.e. Dirichlet data, on their boundary see
for instance,  
\cite{poulton01}, in the context of electromagnetic waves or
\cite{antonakakis14} for clamped inclusions in full vector elasticity. 
 One can anticipate this, following \cite{poulton01} any Floquet-Bloch
eigenvalue problem involving inclusions with zero Dirichlet data on
part of their boundary, will display zero frequency stop-bands, since
the existence of an eigenfield associated with a zero frequency would
imply by the maximum principle that the eigenfield is null
everywhere. Therefore, any wave (electromagnetic, acoustic,
hydrodynamic, elastodynamic) will be prohibited to propagate within a
periodic structure at very low frequencies provided inclusions can be
modelled with zero Dirichlet data (e.g. infinite conducting inclusions
in the case of electromagnetics and clamped inclusions in the case of
elastodynamics).

Although we have concentrated here on the lowest dispersion curve, at
high frequencies, the dispersion curves and asymptotics based around
generating dynamic effective media \cite{antonakakis14,auriault2012,boutin2014}  suggest that 
interesting features will also occur. These include 
ultra-directivity whereby a periodically pinned plate behaves like an
extremely anisotropy effective medium at frequencies associated with
flat bands on dispersion diagrams \cite{lefebvre17a}. 
This feature of a flat band can be seen in
Fig. \ref{fig:KLandElastic}(a), for both the Kirchhoff-Love thin plate
limit and the full vector elastic model. Other interesting features
seen in Fig. \ref{fig:KLandElastic}(a) include coalescing dispersion
curves and inflection points (the latter being associated with
hyperbolic-type effective properties \cite{lefebvre17a}). 
Interestingly for the Rayleigh-like
waves that occur in thick plates with unmovable (clamped) inclusions, one expects
similar features to occur, and indeed one can see in
Fig. \ref{fig:idealised} that the dispersion curves display flat bands and
inflection points and so similar interesting and useful effects will
occur at higher frequencies.


Finally, we note that we assumed a linear elastic model for soil and
columns of concrete, however, the former has visco-elastic features,
that could also be implemented e.g. via a Kelvin-Voigt contact
condition at the soil-column interfaces. Intuitively, we expect that
such a model would reduce the local resonances, and the extremely high damping seen in Fig. 5 and 8 around certain frequencies, would be less pronounced.

\section*{Acknowledgements}
R.C. thanks the EPSRC for their support through research grants
EP/I018948/1, EP/L024926/1, EP/J009636/1. S.G. and Y.A. are thankful for an ERC
starting grant (ANAMORPHISM) that facilitated the collaboration with
Imperial College London. 

We also thank A. Diatta for his help in the implementation of
Cartesian PMLs in elasticity.

\section*{References}

\bibliographystyle{unsrt}

\appendix

\section{Summary of material values and parameters}

For clarity and convenience we summarize here the choices of material
and parameters chosen in the text. 

\begin{tabular}{|l|l|l|c|r|}
  \hline
parameters & soil &  bedrock & steel columns & steel plates \\
\hline
depth  &  15 m & 5 m  & $15$ m   &  \\
\hline
Square pitch  &  & $$  & 2 m &  $$ \\
\hline
Honeycomb pitch &  & $$  & 2$\sqrt 3$ m &  $$ \\
\hline
Plate thickness &  & $$  & &  $0.2$ m \\
\hline
column radius &  & $$  & $0.15$ m (Figs. 3b, 4,5bd) &  $$  \\
 &  & $$  & $0.3$ m (Figs. 6b,7b,8b,10,11,12bce) &  $$  \\
 &  & $$  & $0.6$ m (Figs. 6a,12a) &  $$  \\
\hline
Young modulus $E$ (GPa) & 0.153 & $30$  & $200$  &   $200$ \\
\hline
Poisson's ratio ${\nu}$& $0.3$ & $0.3$  & $0.33$  &   $0.33$ \\
\hline
Density ${\rho}$ (kg/$m^3$) & $1800$ & $2500$  & $7850$  &   $7850$ \\
\hline
\end{tabular}

\end{document}